# SOCIAL AND COLLABORATIVE SERVICES FOR ORGANIZATIONS: BACK TO REQUIREMENTS

*Mayla Alimam, Emmanuel Bertin & Noel Crespi*

**ABSTRACT**
Social and collaborative services have widely spread within the enterprises as they play a part in improving productivity and business outcomes. However, the deployment of these services fluctuates between success and failure. This paper intends to assess their deployment and how they can contribute to value creation in different industries. We investigate the relationship between the services' functionalities and the organizational requirement of these services represented by the coordination. We also consider the organizational transformation driven by servitization and emphasize its impact on the act of coordination. We highlight the tight correlation between the functionalities and the requirement in organic forms which suggests a successful deployment in such enterprises. We nonetheless find that, when the servitization is a strategic intent in organizations with mechanistic characteristics, deploying social and collaborative services can contribute to achieving this aim.

## 1. INTRODUCTION

IT services within the enterprises are enabling technologies applied to facilitate the execution of the enterprise's business activities and hence, increase their performance and improve their productivity. Traditionally, these services have been the Information Systems (IS), being delivered in different models such as intranet, Enterprise Resource Planning (ERP) and Electronic Data Interchange (EDI). A continuous evolution of these services is yet observed and new products labelled as "social" or "web 2.0" have appeared and swept through the popular culture to reach the industries. From Computer Supported Collaborative Work tools to Enterprise 2.0 to Enterprise Social software, these social and collaborative services are increasingly being considered in organizations' strategic IT roadmaps. They are attracting enterprises' top management as they contribute to value creation. Based on a survey covering 4,200 companies around the world, the McKinsey Global Institute outlines that 70% of these companies are deploying social and collaborative services (Chui et al. 2012). However, this deployment is not been adequately successful. Only 3% of the companies are declared as fully benefiting of these services.

Researches addressing social and collaborative services in the enterprises are still at an early age. They often focus on their deployment while defining guidelines and outlining factors that promote their adoption (McAfee 2013, Buregio et al. 2015, Yang et al. 2013, Komarov et al. 2014). Even though some studies characterize the social and collaborative services, they fall within the scope of business activities (Williams and Schubert 2011, Kuettner et al. 2013). They assess their impact on enterprises by examining their context of use, which supports different types of activities. A complementary focus is however necessary to take account of the actual organizational needs for such services, along with the organizational transformation trend. Indeed, today's companies are evolving (Daft 2012), and market analysis is highlighting the current servitization process especially in technology driven industries (Marks et al. 2011). As different types of enterprises involve different ways of working, deploying social and collaboration services will, in its turn, differ in accordance with enterprise types.

This paper investigates the relationship between the nature of social and collaborative services and the organizations' characteristics for the purpose of assessing the deployment of these services in different organizations. Our methodology consists of the following: on the one hand, by looking into how these services have evolved over the years, we identify several distinctive functionalities that make them uniquely powerful and urge to their adoption in the enterprises. On the other hand, we examine the basic requirements of such services in organizations in order to identify the key organizational need that requires to be aligned with these services. In addition, we consider the evolution of the organizational forms and how organizations are migrating from traditional, mechanistic and product-centric business towards more organic forms such as service-oriented enterprises and ecosystems. We emphasize the impact of this organizational transformation on their requirement. Finally, we examine the correlation between the services' functionalities and the organizational requirement in different forms. The rest of the paper is structured as follows: Part 2 explores the functionalities of social and collaborative services. Part 3 examines the basic requirements of such services in the organizations. Part 4 describes the relationship between the services and the organizations. Part 5 gives the conclusions.

## 2. EXPLORING SOCIAL AND COLLABORATIVE SERVICES

A large variety of terms have been coined until present to describe the social and collaborative services in the organizations. Some of them were equivalents to others representing shared concepts. Some others introduced new capabilities with their emergence. We trace the evolution of these services over the years while deriving the main functionalities that they are offering.

### 2.1 Evolution of Social and Collaborative Services

Social and collaborative services can be either tools comprising one or several applications, potentially integrated within the enterprise's system in use (e.g. integrated instant messaging and email services as an independent application or an accessible service through the employee's portal), or otherwise, full-feature platforms that accumulate several services in one single place (e.g. an enterprise social network).

The 1980s and 1990s witnessed a conflicting use of terminology including *Groupware*, *Group Support Systems* (CSS, previously group decision support systems) and *Computer Supported Collaborative/Cooperative Work* (CSCW) to describe the study of technology and coordination of groups with multi-disciplinary perspectives. CSCW was the term held by the academic community for an annual CSCW conference since 1986 whereas groupware continued to appear in both CSS and CSCW literatures (Allen 2004).

The term *Social Software* substituted Groupware (Shirky 2003) and got into more common usage in the 2000s (Grudin 1994). However, its popularity is probably related to the convergence of technological, cultural and societal factors that has brought the desire and the ability for people to use technology to communicate, connect and collaborate (Cook 2008) with the aim of developing new social connections and earning reputation in communities (Farnham et al. 2004).

The emergence of Web 2.0 in 2004 and the similarities of its concepts with the social software led to the introduction of *Enterprise 2.0* (E2.0) which was defined as "the use of emergent social software platforms within companies, or between companies and their partners or customers" (McAfee 2006). This latter, entitled *Enterprise Social Network (ESN)* or *Enterprise Social System (ESS)*, penetrated the enterprises as online social networks for employees' professional exchange (Mathiesen and Fielt 2013).

### 2.2 Main Functionalities of Current Services

Through the evolution presented in the above section, we observe different standpoints regarding the functionalities of social and collaborative services. We agglomerate features that have been outlined in previous seminal researches into a single list representing the main functionalities of current services.

From the perspective of CSCW, Grudin and Poltrock (Grudin and Poltrock 2013) introduce a framework categorizing the features they observed in developed CSCW technologies. They identify three variables describing the technologies' behaviour and evaluate each of the variables according to temporal determinants, i.e. whether being practiced on real time basis or asynchronously. Their proposed behaviour categories are: communication (e.g. emails), information sharing (e.g. document management systems) and collaborative coordination (e.g. group calendars and social networks).

From the perspective of E2.0, the technologies are built into platforms linking together collaborators by new ways of communicating. The communication processes are performed either through channels, i.e. person to person / persons, or following a publish/subscribe pattern (Eugster et al. 2003) over communities on a larger public scale. Since he has coined the term E2.0, Andrew McAfee is continuously emphasizing, in his work, the practice of knowledge management and information exchange that is ensured by these forms of communication processes (McAfee 2006, McAfee 2013). Moreover, current services such as ESSs, surpass their predecessors by enabling a social interaction between the users while controlling their rights and distributing the access, e.g. controlling the ability to create, modify or access a content through the service (Chui et al. 2012).

Bringing together all the mentioned functionalities delivers the following list: communication, supporting knowledge management which includes information sharing and organizing, collaborative coordination, supporting teams represented by the social interaction between group or community members, supporting social behaviours and providing the ability to build a social network of strong and weak ties, supporting different levels of access and controlling rights. This list is summarized in Table 1.

| *Functionalities of Services* |
|---|
| Communication |
| Knowledge management |
| Collaborative coordination |
| Team support |
| Social behaviour |
| Controlled access and rights |

Table 1: Main functionalities of social and collaborative services

## 3. ORGANIZATIONAL REQUIREMENTS FOR DEPLOYING SOCIAL AND COLLABORATIVE SERVICES

The various functionalities we introduced above are related differently to the organizations according to the organizational form of each enterprise.

### 3.1 Evolution in Organizations: Towards Servitization

Standing as a mean to accomplish the organization's overall goals, organization's forms are indeed not all similar. They vary in accordance with the organization's management and strategy. In fact, they balance between two edges: the mechanistic edge and the organic edge (Burns and Stalker 1972, Daft 2012). On the one hand, the mechanistic model represents the rigidity and the verticality of strictly formalized business processes, as for instance in Fordist or bureaucratic organizations. On the other hand, the organic model represents the fluidity and the horizontality (Ylinen et al. 2014). The latter model is increasingly becoming a goal for companies searching to face sustainable challenges such as the organization's growing size and the competitive changeable environment outside its boundaries (Daft 2012). The transformation between the two edges is led in manufacturing firms by servitization, driving the company towards service provision (Vendrell-Herrero et al. 2014). It engenders new ways of working that require being investigated and equipped with the appropriate IT services (Baines and Howard 2013). In both models, the major driver for social and collaborative services consists in supporting and enhancing the coordination.

### 3.2 Coordination: the Key Requirement in Organizations

Coordination is described in the literature as the linkage between parts of an organization (Van de Ven et al. 1976). It represents the core of collaborative work and the driver for achieving the organizations' activities and processes (Okhuysen 2009). Consequently, equipping this key requirement with powerful IT services such as the social and collaborative services may ensure a better accomplishment of the interdependent tasks. Nevertheless, the form of this coordination diverges considerably in organizations. In fact, it is tightly related to the dominant organizational structure.

Mechanistic organizations imply a high formalization and respect of standard procedure. In a pre-defined manner, information flows vertically up the organization's hierarchy and business processes get broken down into strict tasks (Daft 2012).

Moving towards organic forms such as service-oriented companies and ecosystems, coordination's complexity strongly rises. Employee empowerment (Jiang et al. 2011) and the new ways of working occurring in these forms cause the information to flow cross-departmentally and in all directions. This results in a large amount of information to be handled and might therefore get challenged by an information overcharge.

### 4. RELATIONSHIP BETWEEN THE SOCIAL AND COLLABORATIVE SERVICES AND THE ORGANIZATIONS

Having identified the coordination practices in an organization as its key requirement for deploying social and collaborative services, we illustrate in this section the relationship between this requirement on the one hand, and the services represented by their main functionalities on the other hand. Figure 1 demonstrates this relationship. As shown in the figure, the organizational evolution is illustrated on a continuum ranging from the mechanistic to the organic edge. The servitization process represents the evolution axis towards organicity.

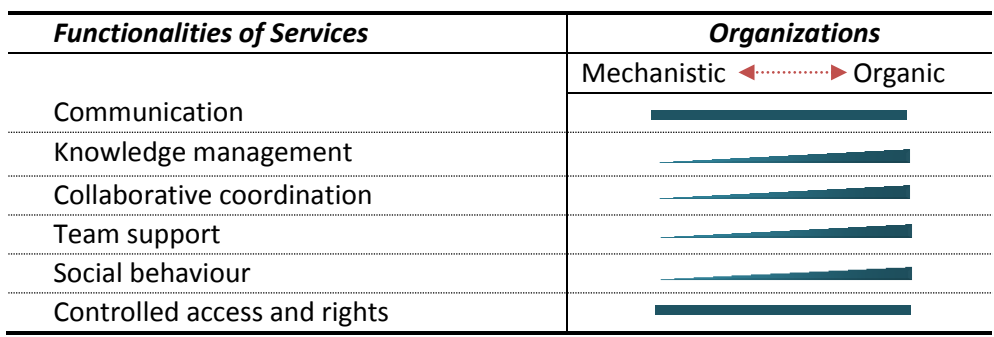

Figure 1: The relationship between the functionalities of social and collaborative services and the organizational key requirement

In fact, the organizational need for each of the services' functionality is directly linked to the coordination practices in a particular company. Thus, as the coordination's complexity levels for the two extreme models are divergent, the need of functionalities might change. This latter is illustrated in the Figure 1 for each functionality. For instance, communication is highly important in both mechanistic and organic models since information flow, whether vertical or horizontal, circulate through communication. On the other hand, the need for team support is less important in mechanistic models than in organic ones. This is because of the horizontal cross-departmentally coordination that is highly required in organic forms.

However, organizations are a hybridization of characteristics of each of the two edges only with different degrees. They therefore have different degrees of needs depending on their position on the continuum. Our findings suggest a high correlation between the services' functionalities and the organizations on the organic edge. We relate this to their close fit to the way of working in enterprises with higher organic characteristics such as service-oriented companies. Our findings also suggest a significant decline of correlation in organizations with more mechanistic characteristic.

These findings underline the relationship between the services and different types of companies. They suggest that deploying these services in organic companies leads to a successful adoption and assists, on the other hand, the transformation processes in evolving ones. Giving as an example the Zero email program announced by Atos in 2011[1]. Gartner's research note described that program as "a big change effort" supporting established ways of working and behaviors (Bradley and Searle 2014).

Nevertheless, challenges might arise to influence this success. Such challenges are the organization's size, user acceptance of these services, the technology already being in use and its overload effect.

## 5. CONCLUSION

Describing the relationship between the social and collaborative services and different forms of organization legitimizes the lack of their deployment's success in some companies. We highlight the tight correlation between the services and organicity driven by servitization. Despite some remaining challenges, the adoption of these services in an organic company should not fail. It shall lead the company to a better productivity and contribute to its value creation.

Nonetheless, we find that deploying such services in an organization with mechanistic characteristics assists the company along its journey towards servitization. It tackles some of the barriers that are challenging the servitization process as for instance, the organization's architecture and capabilities (Nudurupati et al. 2013).

To carry out this work, we suggest performing a complete requirement analysis comprising a detailed study of coordination practices in distinctive organizations. Surveys about the services implication as well as stakeholder interviews can serve as a feedback to support this academic research in order to validate the proposed requirement list. Furthermore, an analysis of the features of different types of services is to be carried out to identify the specifications list that will allow prototyping future services.

---

[1] Atos is an international Information Technology services company headquartered in France. Its initiated Zero email program aims to completely eliminate the internal email exchanges through the use of alternative communications across its collaborative platform "Bluekiwi".

**AUTHOR CONTACT DETAILS**

Mayla Alimam
Orange Labs, mayla.alimam@orange.com

Emmanuel Bertin
Orange Labs, emmanuel.bertin@orange.com

Noel Crespi
Telecom SudParis, noel.crespi@telecom-sudparis.eu